# Dynamic Modeling of Competing Technology Designs, Pricing and Consumer Dynamics

**Philip Vos Fellman**
Southern New Hampshire University
Manchester, NH
Shirogitsune99@yahoo.com

**Adam Groothuis**
Harvard-M.I.T. Division of Health Sciences
Cambridge, MA
Groothua@aol.com

**Sharon Mertz**
Research Director
Gartner, Inc.
Nashua, NH
sharon.mertz@gartner.com

**Roxana Wright**
Keene State College,
Keene, NH
rox_wright@yahoo.com

## Introduction

The Windrum-Birchenhall model of technology succession approaches the problem of competing technologies by developing an n-agent simulation of consumer behavior where two technologies compete for market space. In order to model market dynamics, they express preferences in terms of production costs, price and performance quality which they formalize as a probability of a user adopting new technology B rather than existing technology A at time T expressed as:

$$\Pr \{ U_A(x_A) + v_A(m-p_A) < U_B(x_B) + v_B(m-p_B) \}$$

where $x$ is the characteristic vector of a technology design
$p$ is the price of that design
$V$ is the indirect utility of money that can be obtained in other markets

assuming all other markets are fixed, and that the above function has a constant form.

Our difficulty in expanding this model began with the fact that competing technologies were limited to two, and that coefficients for various inputs to the model, especially design components appeared to be rather arbitrary. While the model was extremely useful in simulating consumer behavior, it was less useful in simulating firm behavior, particularly where there were multiple competing designs.

## The SNHU Model of Technology Succession

The model which we developed allows for five firms with variations in any and all characteristics. Where Windrum and Birchenhall simplified the problem by limiting the simulation to essentially two firms with n-adopters, we have approached the problem by limiting the simulation to five firms, but in an n-factor simulation.

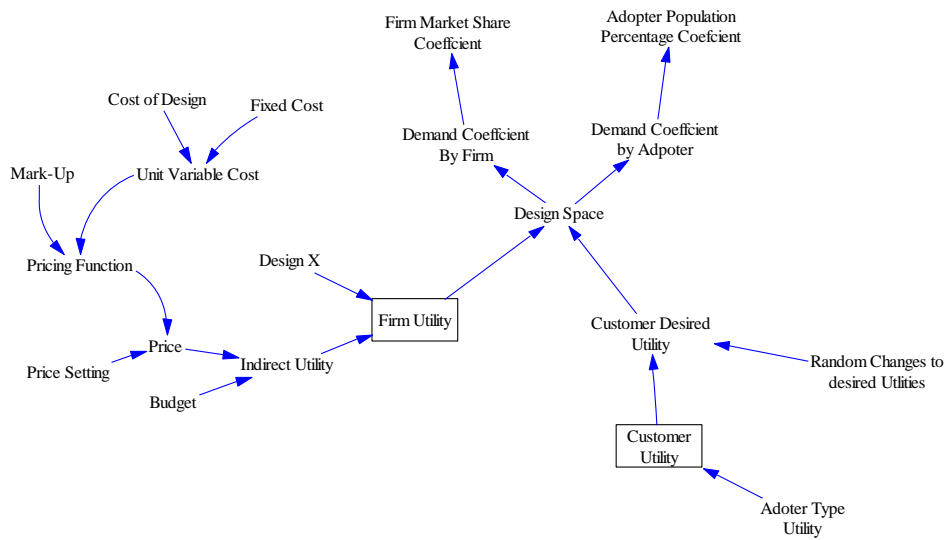

**Figure 1: The SNHU Technology Adoption Model**

Once we establish the basic parameters for the firms, the model allows for interactions of these firms with three simultaneous consumer groups which we also characterize as adopters or adopter groups (to allow for modeling enterprise adoption as well as consumer adoption).

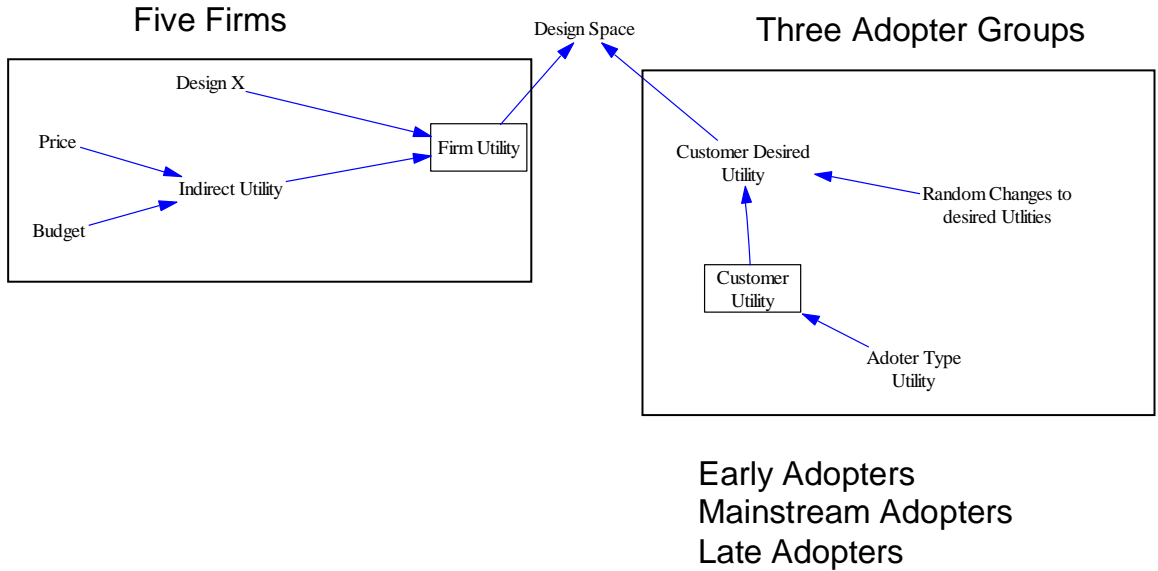

**Figure 2: Firms and Adopters**

The consumer dynamics in the model are based on calculating consumer utilities with random variations in desired utility based on relative adopter utility. The firm and adopter coefficients simulate market adjustments to changes in firm offerings and adopter utility.

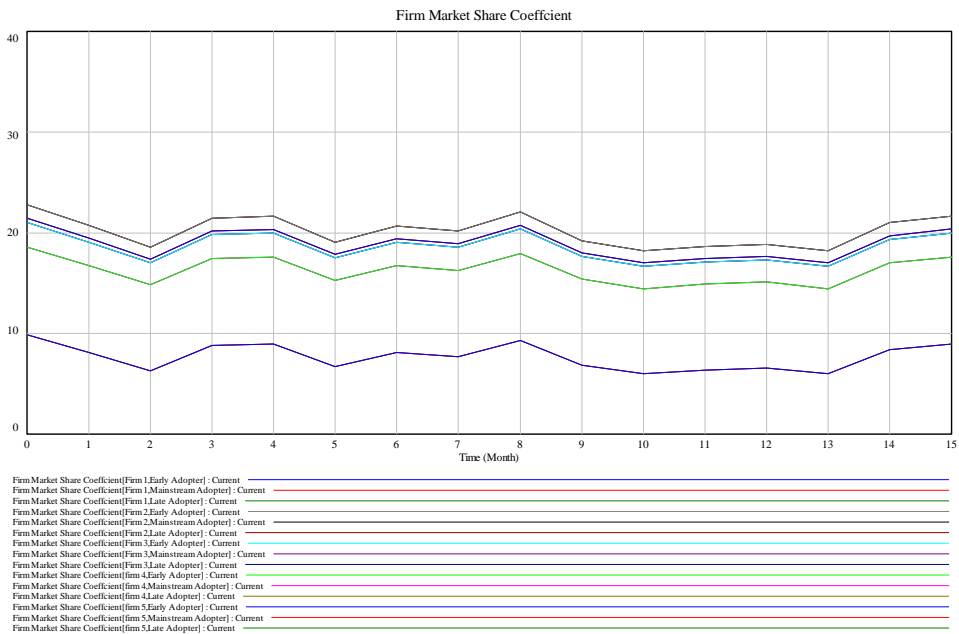

**Figure 3: Market Share by Firm**

The above chart shows the baseline output of the model, in terms of market share after firms have offered their technologies. The model output is the result of the dynamic interaction of consumer utilities and firm offerings.[1] The core of the SNHU model, and the most significant difference between our simulation and that of Windrum and Birchenhall is our treatment of the design space. In order to expand upon their simplified square root function (which is a somewhat arbitrary nonlinear design function) we combine the original model of design space with a ratio of firm utility to customer desired utility creating a new landscape of interaction.

In the next tier of the model, demand coefficient by firm is calculated as the summation of adopter and customer utilities for each respective firm (shown below).

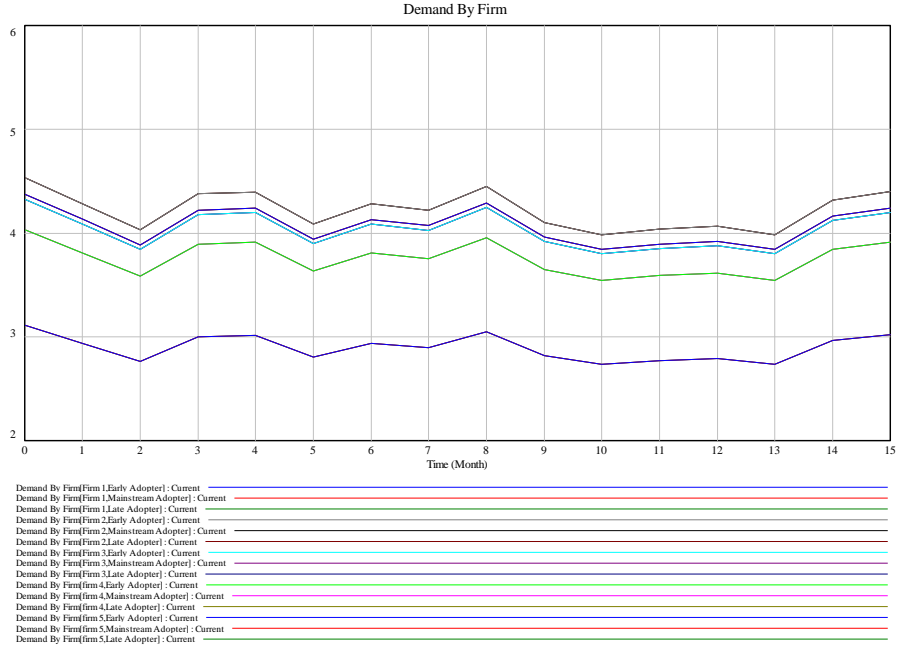

**Figure 4: Demand Coefficient by Firm**[2]

---

[1] In the Windrum-Birchenhall model there were M user groups. Associated with group i = 1…M, a utility function $u_i$ is defined over the offer space, namely the Cartesian product X·P of design space X and the price space P (positive real numbers) of the form:

$$U_i(x,p) = \sum_k \alpha_{ik} v_k(x_k) + \beta_i w(m - p) = \alpha_i \cdot v(x) + \beta_i w(m - p)$$

Here *m* is the budget of the user and is assumed to be the same for all users. The term
$\beta_i w(m - p)$ is the indirect utility obtained by spending the residual budget in other markets. All users in the same group are assumed to adopt the same utility function. Each supplier offers to sell a good with some design *x* at some price *p*. Users use these utility functions to rank alternative offers and as a measure of well-being. Note that users always have the option of not accepting any of the offers and may keep all of their budget for use elsewhere. The utility of this option is $\beta_i w(m)$ and will be called the null utility. It can be seen that the utility functions differ across groups only in having different values for the coefficients $\alpha_i$ and $\beta_i$." (p.5)

In our model, we have maintained the integrity of the firm utility equation, as well as the consumer budget. However, the consumers now have a function called "desired utility" and the adopters have a function called "adopter type utility" each of which rank orders the desirability of the firm offerings.

[2] Interpretation of this chart is a bit complex as it represents fifteen simultaneous variables. The top line in the chart is the summation of the coefficients of the adopters for that firm.

Utility is this context is slightly different than the conventional economic notions of utility. Like the Windrum-Birchenhall model, a lower firm utility coefficient in our model does not necessarily mean a less desirable offering to the consumer. Customer preferences are only expressed when the customer desired utility interacts with the firm utility in the design space. The results of this interaction, in terms of firm market share are shown in figure five.

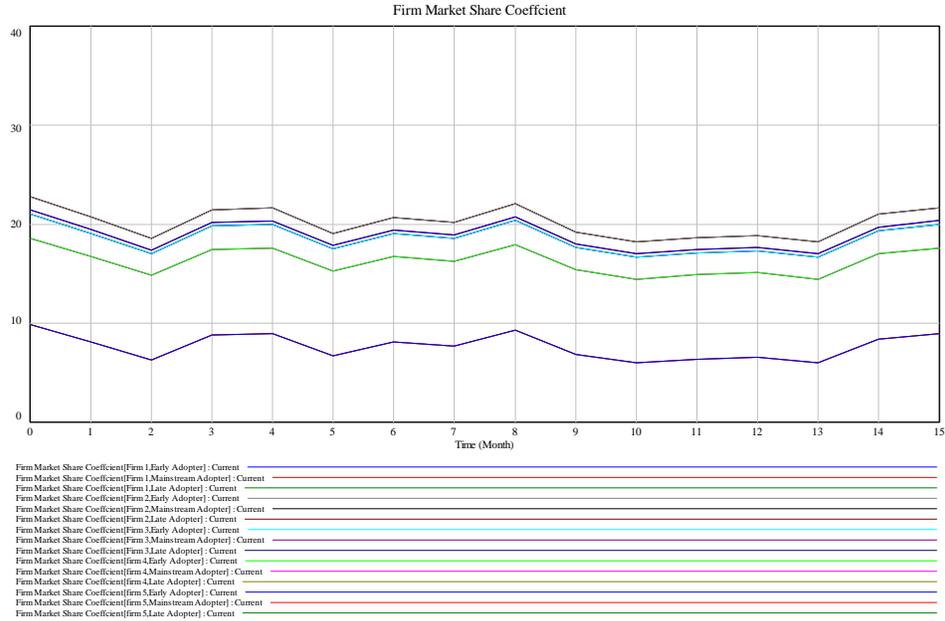

**Figure 5: Firm Market Share Coefficients**

## Technology Succession

Firm five has been singled out here in order to demonstrate a technology succession. What we did in this particular simulation run was to change both the price setting of firm five's offering as well as the design of the product. In essence, firm five was able to create a new product at a competitive price with existing products. In the initial simulation, firm five offered a specialized product at a comparatively high price, representing the "high end" of the market. After the adjustment, firm five is able to offer a new technology exceeding all previous product functionality and is able to offer this product at a competitive price (i.e., a price which matches customer desired utility). The impact on market share coefficient is dramatic (see figure six).

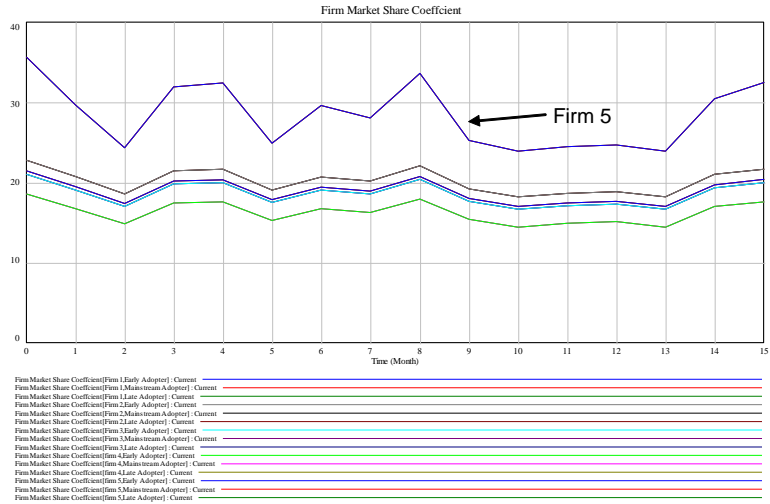

**Figure 6: Firm Market Share Coefficient**

The results from the initial testing of the model's base framework can be extrapolated to reflect the impact of economic shocks and market shifts induced by supplier actions. Continued testing of the model indicates observable trends driven by price/design ratios and impacted by both externalities and agent behaviors. We tested the base model to see if we could mimic both macro and micro level economic effects in order to more appropriately customize the model for specific industry firms and respective consumer groups.

## Model Testing: Macro Economic Effects

To mimic an economic shock we reduced all consumer budgets equally across all firms. The firms with the most unfavorable price to design ratio either fell out of the model entirely or became subject to reduced market performance. Firms that fell out of the model typically represented companies selling products below cost. Further budget reductions were introduced only for firms with unfavorable design to price ratios. The results showed that such firms price themselves out of the market under conditions of an economic shock. A third scenario was tested for firms that had higher design functionality as well as higher associated prices. The budgets for these firms were increased and resulted in a market-leader shift. Essentially, customers at the high end of the market were able to accommodate pricing shifts in order to preserve high-end features. Given differing economic conditions the SNHU model is able to mimic macroeconomic effects resulting in various shifts of market leaders depending upon model variables.

## Model Testing: Micro level Effects

The current model has firms in direct competition. Shifts in the price setting of one firm to gain market share via price sponsorship is captured within the model (figure 4). The model is able to capture these results and simulate consumer reactions to new competitive strategies, intra-firm design changes (something which the Windrum-Birchenhall model is not able to simulate), and changes in internal price-setting.

## Conclusion

While we are still in the early stages of model development, we have been able to generate a number of interesting simulation results. We have modeled technology shocks, new technology entries and what we believe is perhaps the first simulation of firm entry and exit in highly volatile, short product life-cycle high-technology markets.[3] Interestingly, the simulation accurately models the recent battle for market share of the drug eluting stent market between Taxus and Cypher replacing the entire previous bare metal stent technology. In future simulations we expect to further explore the underlying dynamics of pricing and market volatility, particularly for the software industry.

---

[3] See Groothuis, S., Mertz, S. and Fellman, P. (2006) "Agent based simulation of punctuated technology succession : a real-world example using medical devices in the field of interventional cardiology", paper to be presented at the 6th International Conference on Complex Systems, Quincy, MA June 2006.

# Appendix: Technology Adoption under conditions of increasing, decreasing and stable marginal returns to heterogeneous adopters[i]

The model discussed below was developed by Brian Arthur in order to consider what happens when different technologies compete for market share under the differing scenarios of increasing, decreasing and stable marginal returns for adopters (i.e. a network externality effect as opposed to an economy of scale production effect).

In the case of heterogeneous adopters[ii], the model takes into account the payoffs of two unsponsored technologies competing for the replacement of an obsolete technology. For two equal categories of agents differing in their preferences (R agents have a natural preference for technology A, and S agents have a preference for technology B), the variants of A and B available for choice change with the numbers $n_A$ and $n_B$ of previous adoptions. From the observer's point of view, the sequence in which agents make their choices is unknown, in that R and S agents have an equal likelihood of standing in the nth position, while the return functions are known and the demand for one agent is inelastic. Arthur models these returns as shown in the table below:

|         | *Technology A* | *Technology B* |
|---------|----------------|----------------|
| **R Agent** | $a_R + rn_A$ | $b_R + rn_B$ |
| **S Agent** | $a_S + sn_A$ | $b_S + sn_B$ |

The allocation is described by the sequence $\{x_n\}$, where $x_n$ represents the market share of technology A after n choices have been made. The characteristics of the process can be defined as follows:

The process is **predictable** if there is an ex-ante forecasting sequence $\{x_{n'}\}$ such that $|x_{n'} - x_n| \to 0$ with probability 1 as $n \to \infty$ (the observer can accurately predetermine market share as the initial fluctuations average away). The process is held to be **ergodic** if, given two samples of possible historical events $\{t_i\}$ and $\{t_i'\}$ with time paths $\{x_n\}$ and $\{x_{n'}\}$, then $|x_{n'} - x_n| \to 0$ with probability 1 as $n \to \infty$, which implies that different sequences of historical events lead to the same outcome.

The process is **path efficient** if, for $k \leq j \leq m$, $\Pi_\alpha(m) \geq \text{Max}_j \{\Pi_\beta(j)\}$ at any time n, where a more adopted technology $\alpha$ at variant m has the payoff $\Pi_\alpha(m)$, and $\beta$ is a less adopted technology. The above expression stands for the fact that at any time, the development of the less-adopted technology would not have had a higher payoff.

In the situation characterized by constant marginal returns, if $n_A(n) - n_B(n) = d_n$ is the difference between the number of selections of A and B after n choices, the market share of A is $x_n = 0.5 + d_n/2n$, an expression that fully describes the dynamics of adoption of A versus B (R agents always chose A and S agents always chose B).

In the familiar situation of diminishing marginal returns, R agents will start by choosing technology A, but as returns decrease, they will change their preference to B, if the numbers using A become sufficiently greater than the numbers using B, then:

$n_A(n) - n_B(n) = d_n > \Delta_R = (a_R - b_R)/(-r)$ (similar for S agents). This process follows a random walk with reflecting barriers.

In the case of increasing returns, R agents will switch to B if the adoption of B pushes its payoff ahead of A, while the same process will work in reverse for the S agents. As both agent categories begin to choose the same technology, this technology further increases its lead and becomes locked-in.

According to random walk theory, $d_n$ becomes absorbed with probability one, such that the market share of A has to become zero or one, since the two technologies cannot co-exist under a situation of increasing marginal returns. This result is in stark contrast to the better known case of diminishing marginal returns where, $d_n/2n \to 0$ as $n \to \infty$, and $x_n$ approaches 0.5  Under these conditions there will be a market split between the two technologies. Not surprisingly under the constant marginal returns scenario, the standard deviation of $d_n$ increases with $\sqrt{n}$, so that $d_n/2n$ disappears and $x_n$ tends to zero, showing a predisposition towards equal market shares.

---

[i] This treatment was developed by Roxana Wright from W. Brian Arthur's work, "Increasing Returns and Path Dependence in the Economy", University of Michigan Press, 1994.